\documentstyle[preprint,prd,aps,epsfig,12pt]{revtex}

\begin{document}
\draft
%%%%%%%%%%%%%%%%% End of Preamble %%%%%%%%%%%%%%%%%%%%%%%
%%%% Start of Text %%%%%%%%%%%%%%%%%%%%%%%%%%%%%%%%%%%%%%%%%%%%%%%%%%%%%%%
\preprint{
\vbox{
\halign{&##\hfil\cr
}}
}
\title{\LARGE{Decays of the Meson $B_c$ to a $P-$Wave Charmonium State $\chi_c$ or $h_c$}}
\author{Chao-Hsi Chang$^{a,b}$, Yu-Qi Chen$^{a,b}$ Guo-Li Wang$^{b,c}$ and Hong-Shi Zong$^{b,d}$.}

\address{$^a$ CCAST (World Laboratory), P.O. Box 8730, Beijing 100080, China\footnote{Not post-mail address.}}
\address{$^b$ Institute of Theoretical Physics, Academia Sinica, P.O. Box 2735,
Beijing 100080, China}
\address{$^c$ Department of physics, FuJian Normal University, FuZhou 350007, China}
\address{$^d$ Department of Physics, Nanjing University, Nanjing,
210008, China}
%\large{
\maketitle
\begin{center}
\begin{abstract}
The semileptonic decays,
$B_{c}{\longrightarrow}{\chi_c}(h_c)+{\ell}+{ {\nu}}_{\ell}$, and
the two-body nonleptonic decays,
$B_{c}{\longrightarrow}{\chi_c}(h_c)+h$, (here $\chi_c$ and $h_c$
denote $(c\bar c[^3P_J])$ and $(c\bar c[^1P_1])$ respectively, and
$h$ indicates a meson) were computed. All of the form factors
appearing in the relevant weak-current matrix elements with $B_c$
as its initial state and a $P$-wave charmonium state as its final
state for the decays were precisely formulated in terms of two
independent overlapping-integrations of the wave-functions of
$B_c$ and the $P$-wave charmonium and with proper kinematics
factors being `accompanied'. We found that the decays are quite
sizable, so they may be accessible in Run-II at Tevatron and in
the foreseen future at LHC, particularly, when BTeV and LHCB, the
special detectors for B-physics, are borne in mind. In addition,
we also pointed out that the decays $B_c\to h_c+\cdots$ may
potentially be used as a fresh window to look for the $h_c$
charmonium state, and the cascade decays, $B_c\to
\chi_c[^3P_{1,2}]+l+\nu_l$ ($B_c\to \chi_c[^3P_{1,2}]+h$) with one
of the radiative decays $\chi_c[^3P_{1,2}] \to J/\psi+\gamma$
being followed accordingly, may affect the observations of $B_c$
meson through the decays $B_{c}\to {J/\psi}+{l}+\nu_{l}$ ($B_c\to
J/\psi+h$) substantially.
\end{abstract}
\end{center}


\section{Introduction}
\indent

The meson $B_c$, being a unique meson, contains two different
heavy flavors. It decays by one of the two heavy flavors through
weak interaction and it happens that the two have a comparable
possibility each other in magnitude, or by the two heavy flavor
annihilation, hence, its decay-channels which have a sizable
branching ratio, are manifested much richer than those of the
mesons $B^\pm, B^0, B_s, D^\pm, D^0, D_s$ etc. Therefore one may
study the two heavy flavors $b, c$ simultaneously with the meson
$B_c$ alone, as long as its different weak decay channels can be
distinguished from each other well. Of all the mesons, in studying
two heavy flavor $b, c$ simultaneously, $B_c$ is unique.

The meson $B_c$ is just discovered very recently. The first
positive observation was successful in CDF at Tevatron, Fermilab
through the semi-leptonic decays
$B_{c}{\longrightarrow}{J/\psi}+{l}+\nu_{l}$, and the mass
$m_{B_c} = 6.40\pm 0.39 \pm 0.13$ GeV, the lifetime $\tau_{B_c} =
0.46^{+0.18}_{-0.16} \pm 0.03$ ps etc were obtained\cite{cdf}.

Before the observation of CDF, $B_c$-meson
production\cite{cch01,cch02,cchp,masetti},
spectroscopy\cite{quigg,bere} and various
decays\cite{cchd,dec,ccwz} had been widely computed. Now the
further experimental studies of the meson are planned at Tevatron
(in Run II) and at LHC etc. Particularly, in addition to CDF, D0,
ATLAS and CMS, the detectors BTeV and LHCB are specially designed
for B-physics, numerous $B_c^{\pm}$ events (more than $10^8 \sim
10^{10}$ per year) at these two colliders are expected to be
recorded\cite{cch02,cchp}, so a lot of interesting decay channels
of $B_c$ will be well-studied experimentally, and certain rare
processes will become accessible. Therefore, further extensive
theoretical studies of this meson are freshly motivated.

The semileptonic decays, $B_{c}\to \chi_c (h_c) + {l}+\nu_{l}$,
and the two-body nonleptonic decays, $B_c\to \chi_c(h_c)+h$, i.e.
the decays of the meson $B_c$ to a P-wave chamonium state are
certainly interesting, but still missing in literature, thus we
devote this paper to report our the latest computations on them,
although the semileptonic decays were reported shortly\cite{ccwz}.
Why the decays interest people, let us outline the reasons below.

First of all, people would like to know how sizable the decays
will be, especially, to know if accessible in Run-II of Tevatron
and/or in LHC. Especially the cascade decays of $B_c\to
\chi_c+\cdots$ and $\chi_c\to J/\psi+\gamma$ looks quite like as a
signal for the observation of the meson $B_c$ through the decays
$B_c\to J/\psi+\cdots$, because the photon may be missed in
detectors. In addition, two of the P-wave charmonia have a
branching ratio about a few tenth for the radiative decays
$\chi_c[^3P_{1}] \to J/\psi+\gamma$ ($Br=27.3\%$) and
$\chi_c[^3P_{2}] \to J/\psi+\gamma$ ($Br=13.5\%$), so indeed the
cascade decays may potentially contribute a substantial background
for the observation $B_c$ meson through $B_c\to J/\psi+\cdots$.
Therefore, even only from the point of view to estimate the
background for the observation on $B_c$ meson, to see how great
the concerned decays is very interesting.

If one would like to see $CP$ violations in $B_c$ decays, for
example, to see CP violation in the decays $B_c\to h+h_1+h_2$ ($h,
h_1, h_2$ denote various possible mesons), as emphasized in
Ref.\cite{cp}, one knows that the interference of the direct
decays with a cascade one through a resonance, e.g.,
$\chi_c[^3P_0]$, i.e., $B_c\to \chi_c{[^3P_0]}+h$ and
$\chi_c[^3P_0]\to h_1+h_2$, may enhance the visible $CP$ violation
effects substantially. Thus to see the advantage of this method
for the purpose quantitatively, the knowledge on the decay $B_c
\to \chi_c {[^3P_0]}+h$ is necessary.

QCD-inspired potential model works very well for nonrelativistic
double-heavy systems. The systems ($c\bar{b}$) and ($\bar{c} b$)
in forming bound states, except the reduce mass, are similar to
the well-studied systems ($b \bar{b}$) and ($c \bar{c}$), so it is
believed that with potential model the static properties of the
systems ($c\bar{b}$) and ($\bar{c} b$) can be predicted very well
as those of bottomium $(b \bar{b})$ and charmonium $(c \bar{c})$.
In general, to apply the wave functions to computing the relevant
decay matrix elements is attracting, since the potential model
will have further tests. Thus with the wave functions of $B_c$
(the ground state of the system of $(c\bar b)$) and $\chi_c (h_c)$
(the P-wave states of $(c\bar c)$) obtained by potential model, we
have applied the wave functions to compute the decays $B_c\to
\chi_c(h_c)+\cdots$.

Since the mass of $B_c$ ($m_{B_c}$) is much greater than those of
the $P$-wave charmonia ($m_{\chi_c}$ and $m_{h_c}$), so the
momentum recoil in the concerned decays can be a great (even
relativistic). If one tries to apply the Schr\"odinger wave
functions of nonrelativistic binding systems to computing the
decay processes with such a great (even relativistic) recoil
momentum, one cannot carry out the computation of the decay matrix
element successfully just as done in atomic and nuclear decays by
taking a suitable `reference frame' and then a simple `boosting',
since the recoil in an atomic or nuclear process is much smaller
than that in the present concerned decays. The great momentum
recoil obviously means the velocity between the two CMS of $B_c$
meson and the charmonium state is huge, and the potential wave
functions of the parent and the daughter states, given just in
each CMS respectively, cannot be applied directly just by choosing
a suitable reference frame and simple boosting the wave functions
to the same frame. Thus when applying the wave functions to
calculation of the decays (e.g. the semileptonic decays and most
two-body nonleptonic decays here) with such a great (even
relativistic) momentum recoil, special handling is needed. To deal
with the momentum recoil properly, an approach for the decays from
a nonrelativistic $S$-wave state to another $S$-wave one, the
so-called generalized instantaneous approximation, was proposed in
Ref.\cite{cchd}. Since it is straightforward to extend from a
nonrelativistic $S$-wave state to another $S$-wave one, to the
present case, that the decays are from a nonrelativistic $S$-wave
state to a $P$-wave one for the approach, hence here we do so. The
key points of the approach may be outlined as the three steps:
firstly, to `extend' the potential model, which is based on
Schr\"odinger equation, to the one on Bethe-Salpeter (B.S.)
equation\footnote{For the binding systems, $B_c$ and $\chi_c
(h_c)$, to do the extension is just by means of the original
instantaneous approximation proposed by Salpeter, that can be
found in many text book on quantum field theory e.g. the book
\cite{iz} to `build' the relation between the Schr\"odinger
equations and the relevant B.S. ones.} even for the
non-relativistic binding systems; then, according to Mandelstam
method\cite{man} to formulate the (weak) current matrix element
(an elementary factor for the relevant decays) sandwiched by the
B.S. wave functions of the two bound-state, so that the current
matrix element is written in a fully relativistic formulation;
finally, by making the so-called `generalized instantaneous
approximation' on the fully relativistic matrix element i.e. to
integrate out the `time' component of the relative momentum in the
Mandelstam formulation by a contour integration, and as the final
result, the current matrix element turns out back to be formulated
in terms of proper operators sandwiched by the Schr\"odinger wave
functions of the `original' potential model. Since the weak
current matrix (by means of the Mandelstam method) was formulated
relativistically , so we can be sure that the final formulation
takes the recoil effects into account properly and no new free
parameter is added at all. Besides the great recoil effects are
treated properly, one additional advantage of the approach is that
it has a more solid ground on quantum field theory than that on
the `original' potential models, because the B.S. wave functions
and the Mandelstam formulation have a more solid `ground' on
quantum field theories and they are used as a starting point to
make the generalized instantaneous approximation.

On the other hand, B.S. equation is four-dimensional in space-time
to describe a bound state problem, and there are a few problems
still, such as, how to determine the QCD-inspired four-dimensional
interaction kernel of the equation properly, and what is the
physics meaning of the excitation in its relative-time `freedom'
of the two components etc. In addition, the B.S. equation is
harder than a Schr\"odinger one to solve, even when the
four-dimentional kernel is fixed. Whereas with the generalized
instantaneous approximation, the current matrix elements are
reduced into certain proper operators sandwiched by the potential
model Schr\"odinger wave functions finally, therefore, the
approach, in the meantime to circle the difficulty about treating
the great momentum recoil effects properly, has also kept some of
the advantages of potential model, such as to avoid the difficulty
to solve the B.S. equations etc.

Finally we should note here that in our calculating the two-body
nonleptonic decays of $B_c$ to the $P-$wave $\chi_c$ and $h_c$
states, the so-called factorization assumption and the effective
Lagrangian for four fermions in which the `short-distance' QCD
corrections have been taken into account with OPE (operator
product expansion) and RGM (the renormalization group method), as
done by most authors, are adopted.

The paper is organized as follows: To follow the Introduction in
Section-II, the exclusive semileptonic differential decay rates,
the matrix elements and form factors etc are described. In Section
III, the adopted approach, the so-called generalized instantaneous
approximation, to compute the form factors is illustrated
precisely. In Section IV, the two-body non-leptonic decays of
$B_c$ are formulated with necessary description. Finally in
Section V, numerical results and discussions are presented. The
dependence of the current matrix elements on the form factors, and
the dependence of the form factors on $\xi_1$ and $\xi_2$, the
integrations of the wave function overlapping, are put in
Appendix.

\section{The Exclusive Semileptonic Decays and Relevant Current Matrix Elements}

The $T-$matrix element for the  semileptonic decays $B_c\to
X_{c\bar c}+\ell^{+}+\nu_{\ell}$:
\begin{equation}
T=\frac{G_F}{\sqrt{2}}V_{ij}\overline{u}_{\nu_{\ell}}\gamma_{\mu}
(1-\gamma_5)v_{\ell}<X_{c\bar c}(p', \epsilon)\vert J^{\mu}_{ij}
\vert B_c(p)>\;,
\end{equation}
where $X_{c\bar c}$ denotes $\chi_c$ and $h_c$, $V_{ij}$ is the
Cabibbo-Kobayashi-Maskawa(CKM) matrix element and $J^{\mu}$ is the
charged current responsible for the decays, $p$, $p'$ are the
momenta of initial state $B_c$ and final state $X_{c\bar c}$. Thus
we have:
\begin{equation}
\bar{\sum}\vert T\vert^2=\frac{G^2_{F}}{2}\vert V_{ij}\vert^2
l^{\mu\nu}h_{\mu\nu},
\end{equation}
where $h_{\mu\nu}$ is the hadronic tensor and $l^{\mu\nu}$ the
leptonic tensor. The later $l_{\mu\nu}$ is easy to compute whereas
in general the former $h_{\mu\nu}$ can be written as:

$$ h_{\mu\nu}=-\alpha
g_{\mu\nu}+\beta_{++}(p+p')_{\mu}(p+p')_{\nu}+
\beta_{+-}(p+p')_{\mu}(p-p')_{\nu}+\beta_{-+}(p-p')_{\mu}(p+p')_{\nu}+
$$
\begin{equation}
\beta_{--}(p-p')_{\mu}(p-p')_{\nu}+
i\gamma\epsilon_{\mu\nu\rho\sigma}(p+p')^{\rho}(p-p')^{\sigma}\;,
\end{equation}
and by a straightforward calculation, the differential decay-rate
is obtained accordingly:
\begin{center}
$$\frac{d^3\Gamma}{dxdy}=\vert
V_{ij}\vert^2\frac{G^{2}_{F}M^5}{32\pi^3} \left\{
\alpha\frac{(y-\frac{{m^2_l}}{M^2})}{M^2}
+2\beta_{++}\left[2x(1-\frac{M'^2}{M^2}+y)- 4x^2-y\right.\right.$$
$$\left.\left.+\frac{m^2_{l}}{4M^2}(8x+\frac{4M'^2-m^2_l}{M^2}-3y)\right]
\left.+4(\beta_{+-}+\beta_{-+})\frac{{m^2_l}}{M^2}
(2-4x+y-\frac{2M'^2-m^2_l}{M^2})\right.\right.$$
\begin{equation}\left.+
4\beta_{--}\frac{{m^2_l}}{M^2}(y-\frac{{m^2_l}}{M^2})
-\gamma\left[ y(1-\frac{M'^2}{M^2}-4x+y)+\frac{{m^2_l}}{M^2}
(1-\frac{M'^2}{M^2}+y)\right]\right\},
\end{equation}
\end{center}
where $x\equiv E_{\ell}/M$ and $y\equiv (p-p')^2/M^2$, $M$ is the
mass of $B_c$ meson, $M'$ is the mass of final state $X_{c\bar
c}$. The coefficient functions $\alpha$, $\beta_{++}$, $\gamma$
can be formulated in terms of form factors. Note here that we have
kept the mass of the lepton $m_l$ precisely that is different from
those by N. Isgur et al\cite{dec} and by B. Grinstein et
al\cite{s11}, so the formula here can be applied not only to the
cases of $e$ and $\mu$ semileptonic decays but also to those of
$\tau$-semileptonic decays.

1. If $X_{c\bar c}$ is $h_c([^1P_1])$ state: the vector current
matrix element
\begin{equation}
<X_{c\bar c}(p', \epsilon)\vert V_{\mu}\vert B_c(p)>\equiv
r\epsilon^{*}_{\mu}+ s_{+}(\epsilon^{*}\cdot
p)(p+p')_{\mu}+s_{-}(\epsilon^{*}\cdot p)(p-p')_{\mu},
\end{equation}
and the axial vector current matrix element
\begin{equation}
<X_{c\bar c}(p', \epsilon)\vert A_{\mu}\vert B_c(p)>\equiv iv
\epsilon_{\mu\nu\rho\sigma}\epsilon^{*\nu}(p+p')^{\rho}(p-p')^{\sigma}\;,
\end{equation}
where $p$ and $p'$ are the momenta of $B_c$ and $h_c$
respectively, $\epsilon$ is the polarization vector of $h_c$.

2. If $X_{c\bar c}$ is $\chi_c([^3P_0])$ state: the vector matrix
element vanishes, and the axial vector current
\begin{equation}
<X_{c\bar c}(p')\vert A_{\mu}\vert B_c(p)>\equiv
u_{+}(p+p')_{\mu}+u_{-}(p-p')_{\mu}\;.
\end{equation}

3. If $X_{c\bar c}$ is $\chi_c([^3P_1])$ state:
\begin{equation}
<X_{c\bar c}(p', \epsilon)\vert V_{\mu}\vert B_c(p)>\equiv
l\epsilon^{*}_{\mu}+ c_{+}(\epsilon^{*}\cdot p)(p+p')_{\mu}+
c_{-}(\epsilon^{*}\cdot p)(p-p')_{\mu}\;,
\end{equation}
and
\begin{equation}
<X_{c\bar c}(p', \epsilon)\vert A_{\mu}\vert B_c(p)>\equiv iq
\epsilon_{\mu\nu\rho\sigma}\epsilon^{*\nu}(p+p')^{\rho}(p-p')^{\sigma}\;.
\end{equation}

4. If $X_{c\bar c}$ is $\chi_c([^3P_2])$ state:
\begin{equation}
<X_{c\bar c}(p', \epsilon)\vert V_{\mu}\vert B_c(p)>\equiv ih{+-}
\epsilon_{\mu\nu\rho\sigma}
\epsilon^{*\nu\alpha}p_{\alpha}(p+p')^{\rho}(p-p')^{\sigma}\;,
\end{equation}
and
\begin{equation}
<X_{c\bar c}(p', \epsilon)\vert A_{\mu}\vert B_c(p)>\equiv
k\epsilon^{*}_{\mu\nu}p^{\nu}+
b_{+}(\epsilon^{*}_{\rho\sigma}p^{\rho}p^{\sigma})(p+p')_{\mu}+b_{-}
(\epsilon^{*}_{\rho\sigma}p^{\rho}p^{\sigma})(p-p')_{\mu}.
\end{equation}

The form factors $r, s_+, s_-, v, u_+, u_-, l, c_+, c_-, k, b_+,
b_-$ and $h_{+-}$ are functions of the momentum transfer
$t=(p-p')^2$ and can be calculated precisely. In Ref.\cite{cchd}
we proposed an approach, the generalized instantaneous
approximation, to compute those form factors for the decays of
$B_c$ to an $S$-wave charmonium state $J/\psi$ or $\eta_c$. Now we
are computing the form factors $r, s_+, s_-, \cdots$ appearing in
the decays of $B_c$ to a $P$-wave charmonium state, in fact, the
approach may be used directly, thus the approach is adopted in the
present calculations here.

\section{The So-called Generalized Instantaneous Approach
to the Weak Current Matrix Elements}

To calculate these form factors, the approach developed in
Ref.\cite{cchd} is adopted. Let us outline the approach here for
convenience. According to the Mandelstam formalism\cite{man} when
the considered weak (electromagnetic) current matrix element
involves only one hadron in the initial state and one in final
state respectively, then it may be written down in terms of
Bethe-Salpeter (B.S.) wave functions which describe the hadrons as
bound states exactly:
\begin{equation}
l^{\mu}=i\int\frac{d^{4}q}{(2\pi)^{4}}Tr\left[\overline\chi_{p'}(q')
\Gamma^{\mu}\chi_{p}(q)(\not\!{p_{2}}+m_{2})\right]\;,
\end{equation}
where $\chi_{p}(q)$, $\chi_{p'}(q')$ are the B.S. wave functions
of the initial and final states with the corresponding momenta
$p$, $p'$. Throughout the paper we use $p_1$, $p_2$ denote the
momenta of the quarks in the initial meson $B_c$, and $p'_1$,
$p'_2$ denote the momenta of the quarks in the final meson
$\chi_c$ or $h_c$. For convenience let us introduce further
definition of the relative momentum $q$ (or $q'$):
%\begin{center}
$$p_{1}={\alpha}_{1}p+q, \;\;
{\alpha}_{1}=\frac{m_{1}}{m_{1}+m_{2}}\;;$$
$$p_{2}={\alpha}_{2}p-q, \;\;
{\alpha}_{2}=\frac{m_{2}}{m_{1}+m_{2}}.$$
%\end{center}
$p_1, p_2, m_1$ and $m_2$ are the momenta and masses for the quark
and antiquark respectively. Note that the matrix element of the
current Eq.(12) now is fully relativistic, thus it can be used as
the start `point' to take into account the recoil effects in the
decays no matter how great the recoil moment is in the considered
decay. To prepare in applying the generalized instantaneous
approach for the matrix element, we need to `convert' the
potential model onto the B.S. equation `ground'.

\subsection{The Potential Model and B.S. Equation}

In general, the B.S. equation for the corresponding wave function
$\chi_p(q)$:
\begin{equation}
(\not\!{p_{1}}-m_{1})\chi_{p}(q)(\not\!{p_{2}}+m_{2})=
i\int\frac{d^{4}k}{(2\pi)^{4}}V(p,k,q)\chi_{p}(k),
\end{equation}
where $V(p,k,q)$ is the kernel between the quarks in the bound
state, may describe the relevant quark-antiquark bound state well.
Accordingly the B.S. wave function $\chi_{p}(q)$ should satisfy
the normalization condition:
\begin{equation}
\int\frac{d^{4}qd^{4}q'} {(2\pi)^{4}}Tr\left[\overline\chi_{p}(q)
\frac{\partial}{\partial{p_{0}}}\left[S_{1}^{-1}(p_{1})
S_{2}^{-1}(p_{2})\delta^{4}(q-q')+
V(p,q,q')\right]\chi_{p}(q')\right]=2ip_{0}\;,\label{q3}
\end{equation}
where $S_{1}(p_{1})$ and $S_{2}(p_{2})$ are the propagators of the
relevant particles with masses $m_1$ and $m_2$ respectively.

As pointed out in introduction, the B.S. equation in four
dimension should be reduced to a one in three dimension i.e. the
time-like component momentum should be integrated out (the
instantaneous approximation) with a contour integration as
proposed by Salpeter, especially when the kernel has the property
as follows

$$V(p,k,q) \simeq
V(|\stackrel{\rightarrow}{k}-\stackrel{\rightarrow}{q}|)\;,$$

\noindent to do it is very easy. When one make a contour
integration of the `time' component of the relative momentum on
the whole B.S. equation, then the B.S. equation is deduced
straightforwardly into a three-dimensional equation which just is
a Sch\"odinger equation in momentum representation. Since the
start point of the common potential model is a Sch\"odinger
equation, thus we may convert the potential model onto a ground
based on the B.S. equation in the way with instantaneous approach.

To treat the possible great recoil effects in the decays,
furthermore we need to convert the instantaneous approximation to
a covariant way too, i.e. to divide the relative momentum $q$ into
two parts, $q_{\parallel}$ and $q_{\perp}$, a parallel (time-like)
part and an orthogonal one to $p$, respectively:

$$ q^{\mu}=q^{\mu}_{p\parallel}+q^{\mu}_{p\perp}\;, $$

\noindent
where $q^{\mu}_{p\parallel}\equiv (p\cdot
q/M^{2}_{p})p^{\mu}$ and $q^{\mu}_{p\perp}\equiv
q^{\mu}-q^{\mu}_{p\parallel}$. Correspondingly, we have two
Lorentz invariant variables:

$$q_{p}=\frac{p\cdot q}{M_{p}},\;\;
q_{pT}=\sqrt{q^{2}_{p}-q^{2}}=\sqrt{-q^{2}_{p\perp}}\;.$$

In the rest frame of the initial meson, i.e., $\stackrel{\rightarrow}{p}=0$,
they turn back to
the usual component $q_{0}$ and $|\stackrel{\rightarrow}{q}|$, respectively.

Now the volume element of the relative momentum $k$ can be written
in an invariant form:
\begin{equation}
d^{4}k=dk_{p}k^{2}_{pT}dk_{pT}ds d\phi,
\end{equation}
where $\phi$ is the azimuthal angle, $s=(k_{p}q_{p}-k\cdot
q)/(k_{pT}q_{pT})$. Now the interaction kernel can be rewritten
as:
\begin{equation}
V(|\stackrel{\rightarrow}{k}-\stackrel{\rightarrow}{q}|)=V(k_{p\perp},s,q_{p\perp}).
\end{equation}

Defining:
$$ \varphi_{p}(q^{\mu}_{p\perp}) \equiv
i\int\frac{dq_{p}}{2\pi}\chi_{p}(q^{\mu}_{p\parallel},q^{\mu}_{p\perp}),
$$
\begin{equation}
\eta(q^{\mu}_{p\perp})\equiv\int\frac{k^{2}_{pT}dk_{pT}ds}{(2\pi)^{2}}V(k_{p\perp},
s,q_{p\perp})\varphi_{p}(k^{\mu}_{p\perp}).
\end{equation}
The B.S. equation now can be rewritten as:
\begin{equation}
\chi_{p}(q_{p\parallel},q_{p\perp})=S_{1}(p_{1})\eta(q_{p\perp})S_{2}(p_{2})
\label{q1}
\end{equation}
and the propagators can be decomposed as
\begin{equation}
S_{i}(p_{i})=\frac{\Lambda^{+}_{ip}(q_{p\perp})}{J(i)q_{p}+\alpha_{i}M-\omega_{ip}+i\epsilon}+
\frac{\Lambda^{-}_{ip}(q_{p\perp})}{J(i)q_{p}+\alpha_{i}M+\omega_{ip}-i\epsilon},
\end{equation}
with
\begin{equation}
\omega_{ip}=\sqrt{m_{i}^{2}+q^{2}_{pT}},  \Lambda^{\pm}_{ip}(q_{p\perp})=\frac{1}{2\omega_{ip}}\left[
\frac{\not\!{p}}{M}\omega_{ip}\pm J(i)(m_{i}+{\not\!q}_{p\perp})\right],
\end{equation}
where $i=1, 2$ and $J(i)=(-1)^{i+1}$.
Here $\Lambda^{\pm}_{ip}(q_{p\perp})$ satisfies the relations
\begin{equation}
\Lambda^{+}_{ip}(q_{p\perp})+\Lambda^{-}_{ip}(q_{p\perp})=\frac{\not\!{p}}{M}\;,
\Lambda^{\pm}_{ip}(q_{p\perp})\frac{\not\!{p}}{M}
\Lambda^{\pm}_{ip}(q_{p\perp})=\Lambda^{\pm}_{ip}(q_{p\perp})\;,
\Lambda^{\pm}_{ip}(q_{p\perp})\frac{\not\!{p}}{M}
\Lambda^{\mp}_{ip}(q_{p\perp})=0\;.
\end{equation}
Due to these equations, $ \Lambda^{\pm}$ may be referred as the
$p-$projection operators, while in the rest frame of corresponding
meson, they turn to the energy projection operator.

We define $\varphi^{\pm\pm}_{p}(q_{p\perp})$ as
\begin{equation}
\varphi^{\pm\pm}_{p}(q_{p\perp})\equiv
\Lambda^{\pm}_{1p}(q_{p\perp})\frac{\not\!{p}}{M}\varphi_{p}(q_{p\perp})
\frac{\not\!{p}}{M}
\Lambda^{{\mp}{C}}_{2p}(q_{p\perp})\;,\label{q5}
\end{equation}
where the upper index $C$ denotes the charge conjugation. In our notation,
$\Lambda^{{\pm}{C}}_{2p}(q_{p\perp})\equiv\Lambda^{\mp}_{2p}(q_{p\perp}).$
Integrating over $q_{p}$ on both sides of Eq.(\ref{q1}), we obtain:
$$
(M-\omega_{1p}-\omega_{2p})\varphi^{++}_{p}(q_{p\perp})=
\Lambda^{+}_{1p}(q_{p\perp})\eta_{p}(q_{p\perp})\Lambda^{-C}_{2p}(q_{p\perp});
$$
$$(M+\omega_{1p}+\omega_{2p})\varphi^{--}_{p}(q_{p\perp})=
\Lambda^{-}_{1p}(q_{p\perp})\eta_{p}(q_{p\perp})\Lambda^{+C}_{2p}(q_{p\perp});$$
\begin{equation}
\varphi^{+-}_{p}(q_{p\perp})=\varphi^{-+}_{p}(q_{p\perp})=0.\label{q2}
\end{equation}

The normalization condition of Eq.(\ref{q3}) now becomes

$$ \int\frac{q_T^2dq_T}{(2\pi)^2}tr\left[\overline\varphi^{++}
\frac{{/}\!\!\!
{p}}{M}\varphi^{++}\frac{{/}\!\!\!{p}}{M}-\overline\varphi^{--}
\frac{{/}\!\!\! {p}}{M}\varphi^{--}\frac{{/}\!\!\!
{p}}{M}\right]=2P_0\;. $$

From these equations, one may see that in the weak binding case to
compare with the factor $(M-\omega_{1p}-\omega_{2p})$, the factor
$(M+\omega_{1p}+\omega_{2p})$ is large, so the negative energy
components of the wave functions $\varphi^{--}$ are small. In the
present case, for the heavy quarkonium and $B_c$ meson, it is just
the case, so we ignore the negative energy components of the wave
functions safely at the lowest order approximation.

Neglecting the negative energy components of the wave functions,
the B.S. equation contains the positive component

$$\varphi^{++}_{p}(q_{p\perp})\equiv \Lambda^{+}_{1p}(q_{p\perp})
\frac{\not\!{p}}{M}\varphi_{p}(q_{p\perp}) \frac{\not\!{p}}{M}
\Lambda^{{-}{C}}_{2p}(q_{p\perp})$$

only, and the normalization condition becomes:
$$
\int\frac{q_T^2dq_T}{(2\pi)^2}tr\left[\overline\varphi^{++}
\frac{{/}\!\!\!
{p}}{M}\varphi^{++}\frac{{/}\!\!\!{p}}{M}\right]=2P_0 $$

Now let us consider the wave function $\varphi^{++}$ appearing in
the above equations. We know that the total angular momentum of a
meson is composed from orbital one and spin, furthermore there are
two ways i.e. S-L coupling or j-j coupling to compose the total
angular momentum. Here to consider $P$-wave states of charmonium,
we adopt the way of S-L coupling, i.e. we let the spins of the two
quarks couple into a total spin, which can be either singlet or
triplet, then the total spin couple to the relative orbital
angular momentum, and finally we obtain the total angular
momentum. In this way, the reduced B.S. wave function $\varphi_P$
can be written approximately as:
\begin{equation}
\varphi_{^{1}S_0}(\stackrel{\rightarrow}{q})=
\frac{\not\!{P}+M}{2\sqrt {2}M}\gamma_5
\psi_{n00} (\stackrel{\rightarrow}{q}),\label{q6}
\end{equation}
for $^{1}S_0$ state, and
\begin{equation}
\varphi_{^3S_1}^{\lambda}(\stackrel{\rightarrow}{q})=
\frac{\not\!{P}+M}{2\sqrt {2}M}\not\! {\epsilon}^{\lambda}
\psi_{n00} (\stackrel{\rightarrow}{q}),
\end{equation}
for ${^3}S_1$ state, where ${\epsilon}^{\lambda}$ is the
polarization of this state.  For the $P$-wave $(c\bar c)$ wave
functions:
\begin{equation}
\varphi_{^{1}P_1}(\stackrel{\rightarrow}{q})=
\frac{\not\!{P}+M}{2\sqrt {2}M}\gamma_5
\psi_{n1M_z} (\stackrel{\rightarrow}{q}),\label{q7}
\end{equation}
for $^{1}P_1$, i.e. $h_c$ state, and
\begin{equation}
\varphi_{^3P_J}^{J_z}(\stackrel{\rightarrow}{q})=
\frac{\not\!{P}+M}{2\sqrt {2}M}\not\! {\epsilon}^{\lambda}(S)
\psi_{n1M_z}
(\stackrel{\rightarrow}{q})<1S_z,LM_z|JJ_z>,\label{q8}
\end{equation}
for ${^3}P_J$($J$=0, 1, 2) i.e. $\chi_c$ states, where $\epsilon$
is the polarization vector of total spin, $<1S_z,LM_z|JJ_z>$ is
Clebsch-Gordon coefficients which couple $L$, $S$ to the total
angular momentum $J$. $\psi_{n00}$ and $\psi_{n1M_z}$ are the full
B.S. wave functions.

\subsection{The Radius B.S. Equation in Momentum Space}

To solve the B.S. equation, the key problem is about its radial
component. If we ignore the negative energy contributions, the
reduced B.S. equation Eq.(18) in the rest frame of the meson
center mass system can be written as:
\begin{equation}
\varphi_P(\stackrel{\rightarrow}{q})=\frac
{\Lambda^{+}_{1}(\stackrel{\rightarrow}{q})
\int\frac{d\stackrel{\rightarrow}{k}}{(2\pi)^3}
V(\stackrel{\rightarrow}{k},\stackrel{\rightarrow}{q})
\varphi_P(\stackrel{\rightarrow}{k})
\Lambda^{+}_{2}(\stackrel{\rightarrow}{q})}
{M-\omega_1-\omega_2}.\label{q13}
\end{equation}
In the frame, the energy projection operator:
$$\Lambda^{+}_{1}=\frac{1}{2\omega_1}(\omega_1\gamma_0+\stackrel{\rightarrow}{\gamma}
\cdot\stackrel{\rightarrow}{q}+m_1),$$
$$\Lambda^{+}_{2}=\frac{1}{2\omega_1}(\omega_2\gamma_0-\stackrel{\rightarrow}{\gamma}
\cdot\stackrel{\rightarrow}{q}-m_2),$$

\noindent
where the kernel $V$ acts on
$\varphi(\stackrel{\rightarrow}{q})$ as:
\begin{equation}
V(\stackrel{\rightarrow}{q})\varphi(\stackrel{\rightarrow}{q})=
V_s(\stackrel{\rightarrow}{q})\varphi(\stackrel{\rightarrow}{q})
+V_v(\stackrel{\rightarrow}{q})\gamma_\mu
\varphi(\stackrel{\rightarrow}{q})\gamma^\mu,
\end{equation}
i.e. to correspond to the potential model more precisely, the
interaction kernel can be formally divided into the corresponding
non-perturbative QCD `linear' one, $V_s$ (in scalar nature) and
the corresponding gluon exchange one, $V_v$ (in vector nature).

When substituting Eqs.(\ref{q6},\ref{q7}), the wave functions in
the meson center mass system, to the reduced B.S. equation
Eq.(\ref{q13}), the equation for a spin singlet state $S=0$
becomes:
\begin{equation}
\phi_{S=0}(\stackrel{\rightarrow}{q})=
\frac{1}{4\omega_1\omega_2(M-\omega_1-\omega_2)}
\left\{m_1m_2
\int \left[4V_v(\stackrel{\rightarrow}{q},
\stackrel{\rightarrow}{k})-4V_s(\stackrel{\rightarrow}{q},
\stackrel{\rightarrow}{k})\right]
\phi_{S=0}(\stackrel{\rightarrow}{k})d\stackrel{\rightarrow}{k}\right\},
\end{equation}
where the $\phi_{S=0}(\stackrel{\rightarrow}{q})$ is
$\phi_{n00}(^1S_0)$ or $\phi_{n1M_z}(^1P_1)$. Since square of the
relative momentum ${\stackrel{\rightarrow}{q}}^2$ is small to
compare with quark mass squared in the `double heavy' meson, as a
lowest order approximation, we have ignored such higher terms and
use $\omega_1=m_1$, $\omega_2=m_2$ in numerator.

Now let us factorize out the radial component of the wave function
and its relevant B.S. equation in momentum space from the angular
ones:

$$\psi_{nLM_z}(\stackrel{\rightarrow}{q})=
\phi_{nL}(|\stackrel{\rightarrow}{q}|)Y_{LM_z}(\theta,\varphi ),$$
where $n$ is the principal quantum number, $L$ is the orbital
angular momentum and $M_z$ is the projection of the third
component of $L$, $\phi_{nL}(|\stackrel{\rightarrow}{q}|)$ is the
radial wave function and $Y_{LM_z}(\theta,\phi )$ is the spherical
harmonic function. For the spin singlet states, multiplying
$Y^{*}_{LM_z}(\hat{q})$ to two sides of the reduced B.S. equation
and sum over $M_z$ by using the formula,

$$\frac{4\pi}{2L+1}\sum\limits_{M_z}Y_{LM_z}(\hat{q})Y^{*}_{LM_z}(\hat{k})=
P_L(cos\theta),$$

\noindent
where $\theta$ is the angular between the unit vector
$\hat{q}$ and $\hat{k}$, the radial reduced B.S. equation for
$^1S_0$ state is obtained:
\begin{equation}
\phi_{n0}(|\stackrel{\rightarrow}{q}|)=
\frac{1}{4\omega_1\omega_2(M-\omega_1-\omega_2)}
\left\{m_1m_2
\int \left[4V_v(\stackrel{\rightarrow}{q},
\stackrel{\rightarrow}{k})-4V_s(\stackrel{\rightarrow}{q}, \stackrel{\rightarrow}{k})\right]
\phi_{n0}(|\stackrel{\rightarrow}{k}|)d\stackrel{\rightarrow}{k}\right\}.
\end{equation}
Whereas for $^1P_1$ state:
\begin{equation}
\phi_{n1}(|\stackrel{\rightarrow}{q}|)=
\frac{1}{4\omega_1\omega_2(M-\omega_1-\omega_2)}
\left\{m_1m_2
\int \left[4V_v(\stackrel{\rightarrow}{q},
\stackrel{\rightarrow}{k})-
4V_s(\stackrel{\rightarrow}{q}, \stackrel{\rightarrow}{k})\right]
\phi_{n1}(|\stackrel{\rightarrow}{k}|)cos{\theta}d\stackrel{\rightarrow}{k}\right\},
\end{equation}
where $\phi_{n0}(|\stackrel{\rightarrow}{q}|)$ and
$\phi_{n1}(|\stackrel{\rightarrow}{q}|)$ are the radial parts of
the wave functions.

Similarly, for the spin triplet states $S=1$ we have:

$$\sum\limits_{lm}<1S_z,LM_z|JJ_z>\phi_{S=1}(\stackrel{\rightarrow}{q})=
\sum\limits_{lm}<1S_z,LM_z|JJ_z>
\frac{1}{4\omega_1\omega_2(M-\omega_1-\omega_2)}$$
\begin{equation}
\times\left\{m_1m_2
\int \left[4V_v(\stackrel{\rightarrow}{q},
\stackrel{\rightarrow}{k})-4V_s(\stackrel{\rightarrow}{q}, \stackrel{\rightarrow}{k})\right]
\phi_{S=1}(\stackrel{\rightarrow}{k})d\stackrel{\rightarrow}{k}\right\},
\end{equation}
where the $\phi_{S=1}(\stackrel{\rightarrow}{q})$ is
$\phi_{n00}(^3S_1)$ or $\phi_{n1M_z}(^3P_J)$. Then the equation
for $^3S_1$ is:
\begin{equation}
\phi_{n0}(|\stackrel{\rightarrow}{q}|)=
\frac{1}{4\omega_1\omega_2(M-\omega_1-\omega_2)}
\left\{m_1m_2
\int \left[4V_v(\stackrel{\rightarrow}{q},
\stackrel{\rightarrow}{k})-4V_s(\stackrel{\rightarrow}{q},
 \stackrel{\rightarrow}{k})\right]
\phi_{n0}(|\stackrel{\rightarrow}{k}|)d\stackrel{\rightarrow}{k}\right\};
\end{equation}
and for $^3P_J$:
\begin{equation}
\phi_{n1}(|\stackrel{\rightarrow}{q}|)=
\frac{1}{4\omega_1\omega_2(M-\omega_1-\omega_2)}
\left\{m_1m_2
\int \left[4V_v(\stackrel{\rightarrow}{q},
\stackrel{\rightarrow}{k})-4V_s(\stackrel{\rightarrow}{q},
\stackrel{\rightarrow}{k})\right]\phi_{n1}(|\stackrel{\rightarrow}{k}|)
cos{\theta}d\stackrel{\rightarrow}{k}\right\}.
\end{equation}
The normalization of $\phi_{nL}$ now is read:

$$ \int\frac{q_T^2dq_T}{(2\pi)^2}\left[\frac{m_1m_2}
{\omega_1\omega_2}\phi^2_{nL}(|q_T|)\right]=2M. $$

Under the present further approximation, the three triplet
$P$-wave states $^3P_J$ and the singlet $^1P_1$ as well, are
degenerated. The reason is that we have ignored the `splitting'
interactions at all.

\subsection{The Generalized Instantaneous Approximation}

After neglecting the negative energy component and the `treatment'
above, the weak current matrix elements become as follows:

$$ l^{\mu}=i\int\frac{d^{4}q}{(2\pi)^{4}}
Tr\left[\overline{\eta}(q'_{p'\perp})\frac{{\Lambda}'^{+}_{1}(q'_{p'\perp})}
{q'_{p'}+{\alpha}'_{1}M'-{\omega}'_{1}+i\epsilon}
\Gamma^{\mu}\frac{{\Lambda}^{+}_{1}(q_{p\perp})}
{q_{p}+\alpha_{1}M-\omega_{1}+i\epsilon}\right.$$
\begin{equation}\left.
\times\eta(q_{p\perp})\frac{{\Lambda}^{+}_{2}(q_{p\perp})}
{-q_{p}+\alpha_{2}M-\omega_{2}+i\epsilon}\right]\;,
\end{equation}

The generalized instantaneous approximation, being an extension
for the original one on the B.S. equations suggested by Salpeter,
with the Cauchy's theorem performs a contour integration about the
time-like component $q_P$ in complex plan on the whole current
matrix elements precisely. As the final result, the matrix
elements turn out to become a three dimensional integration about
the space-like components $q_\perp$.

If we choose the contour along the lower half plane, after
completing the contour integration, the current matrix elements
become as follows:

$$ l^{\mu}=\int\frac{d^3q_{\perp}}{(2\pi)^{3}}
Tr\left[\frac{\overline{\eta}(q'_{p'\perp}){\Lambda}'^{+}_{1}(q'_{p'\perp})}
{M'-{\omega}'_{1}-{\omega}'_{2}} \Gamma^{\mu}
\frac{{\Lambda}^{+}_{1}(q_{p\perp})\eta(q_{p\perp})
{\Lambda}^{+}_{2}(q_{p\perp})}
{M-{\omega}_{1}-{\omega}_{2}}\right]\;, $$

This matrix elements can also be written in the frame where the
momentum $q'_{\perp}$ is the integral argument by means of a
suitable Jacobi transformation, i.e.
\begin{equation}
l_{\mu}(r)= \int \frac{q'^{2}_{p'T}dq'_{p'T}ds}
{(2\pi)^{2}}tr\left[ \overline{\varphi}^{++}_{p'}(q'_{p'\perp})
\Gamma_{\mu}\varphi^{++}_{p}(q_{p\perp})\frac{\not\! P'}
{M'}\right]\;.\label{q4}
\end{equation}
The above formula with the argument $q'_{\perp}$ as the integral
argument is more convenient, especially, in the cases when we
calculate the matrix elements involving a $P$-wave state in the
final state.

After performing the calculations on the matrix elements $l_\mu$
precisely, the dependence of the matrix elements on the
overlapping integrations of the initial and the final state wave
functions becomes transparent. So is all the form factors too.

Since there is the so-called spin symmetry for heavy mesons, all
of the form factors for their decays may attributed to one
`universal' function i.e. the Isgur-Wise function\cite{i-w}.
Therefore for the double heavy meson $B_c$ to a $S$-wave
charmonium, at the limiting $m_b >> m_c >> \Lambda_{QCD}$ i.e.
turning to the case of the heavy mesons, the form factors are
attributed to the Isgur-Wise function, and the Isgur-Wise function
is related to an overlapping integration of the wave functions of
$B_c$ and the $S$-wave charmonium with certain kinematics factors
precisely\cite{cchd,spin}. Now at the present case of $B_c$ to a
$P$-wave charmonium, not only due to the spin-symmetry but also
due to the great recoil in the decays, alternatively there are two
independent and `universal' functions, essentially, just two
overlapping integrations of the wave functions of the initial and
final bound states, $\xi_1$ and $\xi_2$, and all of the form
factors are described by the two general functions with proper
kinematics factors precisely.

Since in the present case the initial state is of an $S$-wave and
the final state is of a $P$-wave, so the matrix elements must be
related to two kinds of terms: one is to the integration which
does not depend on the relative momentum $q'_{p'\perp}$ at all,
and the one just on $q'_{p'\perp}$ linearly. Namely all the form
factors appearing in the decays depend on two universal functions
$\xi_1$ and $\xi_2$ only:
$${\epsilon}^{\lambda}(L)\cdot\epsilon_0\xi_1
\equiv\int\frac{d^3q'_{p'\perp}}{(2\pi)^3}
\psi'^{*}_{n1M_z}(q'_{p'T})\psi_{n00}(q_{pT}),$$
\begin{equation}
\epsilon^{\alpha}_{\lambda}(L)\xi_2
\equiv\int\frac{d^3q'_{p'\perp}}{(2\pi)^3}
\psi'^{*}_{n1M_z}(q'_{p'T})\psi_{n00}(q_{pT})
q'^{\alpha}_{p'\perp},\label{q9}
\end{equation}
where
$$\epsilon_0\equiv\frac{p-\frac{p\cdot p'}{M'^2}p'}{\sqrt
{\frac{(p\cdot p')^2}{M'^2}-M^2}}$$
describes the polarization
vector along recoil momentum $\stackrel {\rightarrow}{p'}$,
$\epsilon^{\lambda}(L)$ is the polarization vector of the orbital
angular momentum.

We should note that for the decays from an $S$-wave state to a
$P$-state, the function $\xi_1$ generated in the present approach
is special. Since $\xi_1$ has more direct roots to the momentum
recoil, so it cannot be obtained by boosting the final state wave
function as done in the cases with a small recoil. The reason is
that $\xi_1$ approaches to zero when the momentum recoil vanishes.
Whereas, the function $\xi_2$, as in the cases with a small
recoil, can essentially involve recoil effects just by `boosting'
the final state wave function.

Substituting the B.S. wave functions Eq.(\ref{q6}) and
Eqs.(\ref{q7}-\ref{q8}) into the equation of current matrix
elements and using Eq.(\ref{q9}), the precise formula for the form
factors i.e. the precise dependence of the form factors on $\xi_1$
and $\xi_2$, can be obtained and we put them in the appendices and
the curves of $\xi_1$ and $\xi_2$ obtained by numerical
calculations are shown in a figure. With the functions $\xi_1$,
$\xi_2$ and the form factors, the decay rates of the semileptonic
decays and the spectrum of the charged lepton for the decays can
be obtained by straightforward numerical calculations.

Note that in our calculation on the form factors, we have used the
relations: $$\sum\limits_{\lambda,\lambda'}
\epsilon^{\lambda}_{\mu}(S)\epsilon^{\lambda'}_{\nu}(L)<1\lambda;1\lambda'|00>
=\sqrt{\frac{1}{3}}(g_{\mu\nu}-\frac{p'_{\mu}p'_{\nu}}{M'^2}),$$
$$\sum\limits_{\lambda,\lambda'}
\epsilon^{\lambda}_{\mu}(S)\epsilon^{\lambda'}_{\nu}(L)<1\lambda;1\lambda'|1\lambda''>
=\sqrt{\frac{1}{2}}
\frac{i}{M'}\epsilon_{\mu\nu\alpha\beta}{p'^{\alpha}}\epsilon_{\lambda''}^{\beta}(J),$$
\begin{equation}
\sum\limits_{\lambda,\lambda'}\epsilon^{\lambda}_{\mu}(S)
\epsilon^{\lambda'}_{\nu}(L)<1\lambda;1\lambda'|2\lambda''>
\equiv\epsilon^{\lambda''}_{\mu\nu}(J),
\end{equation}
where $<1S_z;1L_z|JJ_z>$ as previous are C.-G. coefficients. The
polarization vector $\epsilon^{\lambda}_{\mu}(J), J=1$ and the
tensor $\epsilon^{\lambda}_{\mu\nu}(J), J=2$ have the projection
properties:
$$\sum_{\lambda}\epsilon^{\lambda}_{\mu}(J)\epsilon^{\lambda}_{\nu}(J)=
(\frac{p'_{\mu}p'_{\nu}}{M'^2}-g_{\mu\nu})\equiv P'_{\mu\nu},$$
\begin{equation}
\sum\limits_{\lambda}\epsilon^{\lambda}_{\mu\nu}(J)\epsilon^{\lambda}_{\alpha\beta}(J)
=\frac{1}{2}(P'_{\mu\alpha}P'_{\nu\beta}+P'_{\mu\beta}P'_{\nu\alpha})-
\frac{1}{3}P'_{\mu\nu}P'_{\alpha\beta}.
\end{equation}

\section{The Two-body Non-leptonic Decays}
\indent

In this section we outline how the two-body non-leptonic decays
$B_c\to \chi_c(h_c) + h$ (here $h$ denotes a meson) are
calculated. We adopt the factorization assumption on the decay
amplitudes which is widely adopted in estimation of the
non-leptonic decays for various mesons. With the assumption, the
weak current matrix elements appear in the calculations precisely
and they are related to the form factors just obtained in the
previous section. For the non-leptonic decay modes $B_c \to
\chi_c(h_c) +h$ (caused by the decay $b \to c$), the following
effective Lagrangian $L_{eff}$ (QCD corrections are involved) is
responsible:
\begin{eqnarray}
L_{eff}={G_F \over \sqrt{2}} &\big\{
&V_{cb}[c_1(\mu)Q_1^{cb}+c_2(\mu)Q_2^{cb}] +h.c. \big\} \nonumber
\\ &+ &\;penguin\;\; operators\; . \label{heff}
\end{eqnarray}

$G_F$ is the Fermi constant, $V_{ij}$ are CKM matrix elements and
$c_i(\mu)$ are scale-dependent Wilson coefficients. The four-quark
operators $Q_1^{cb}$ and $Q_2^{cb}$ (CKM favoured only) are:
\begin{eqnarray}
Q_1^{cb}&=& \big[V_{ud}^* ~({\bar d}u)_{V-A} +V_{us}^* ~({\bar
s}u)_{V-A}+ V_{cd}^* ~({\bar d}c)_{V-A} +V_{cs}^*~ ({\bar
s}c)_{V-A}\big]~ ({\bar c} b)_{V-A} \nonumber \\ Q_2^{cb}&=&
\big[V_{ud}^*~ ({\bar c}u)_{V-A} ~({\bar d}b)_{V-A} +V_{us}^*
~({\bar c}u)_{V-A}~({\bar s}b)_{V-A} + V_{cd}^*~ ({\bar
c}c)_{V-A}~({\bar d}b)  \nonumber \\ &+& + V_{cs}^* ~({\bar
c}c)_{V-A}~({\bar s}b)\big], \label{q12}
\end{eqnarray}
\noindent where $({\bar q}_1 q_2)_{V-A}$ denotes ${\bar q}_1
\gamma_\mu (1-\gamma_5) q_2$.

Because at this moment we restrict ourselves to consider the
decays in which the coefficients of `penguin' operators in the
effective Lagraingen are small in comparison with the two main
ones $c_1$ and $c_2$, so the contribution from penguin terms is
neglected in the calculations, although in the Ref.\cite{lifet} it
is pointed out that in total decay width the penguin may have
interference with the main ones and can course an increase about
$\%3\sim 4$. Moreover, at this stage we also restrict ourselves
only to consider the decay modes where the weak annihilation
contribution is small due to precise reasons e.g. the helicity
suppression etc\footnote{We will consider the contribution from
penguin and weak annihilation carefully elsewhere.}, namely we
neither take into account the contribution from the weak
annihilation here.

Precisely by means of the factorization assumption, the decay
amplitudes for the non-leptonic decays can be formulated into the
three factors: the so-called leptonic decay constants, which are
defined by the matrix elements: $<0|A_\mu|M(p)>=i f_M p_\mu$ (or
$<0|V_\mu|V(p,\epsilon)>=f_V M_V \epsilon_\mu$); the weak current
matrix elements $<\chi_c|V_\mu(A_\mu)|B_c>$, which are those as
the semileptonic decays; and the relevant coefficients in the
combinations: $a_1=c_1+ \kappa c_2$ and $a_2=c_2+ \kappa c_1$,
here $\kappa=1/N_c$ and $N_c$ is number of color. The coefficients
in the combination $a_1, a_2$ is due to the weak currents being
`Fierz-reordered'. In the numerical calculation later on, we will
choose $a_1=c_1$ and $a_2=c_2$, i.e., we take $\kappa=0$ in the
spirit of the large $N_c$ limit, and QCD correction coefficients
$c_1$ and $c_2$ are computed at the energy scale of $m_b$.

Therefore with the relations between the currents and form factors
obtained as in the semileptonic decays, finally the factorized
amplitudes for the nonleptonic decays can be formulated in terms
of the form factors and the decay constants by definitions:
$<0|A_\mu|M(p)>=i f_M p_\mu$ and $<0|V_\mu|V(p,\epsilon)>=f_V M_V
\epsilon_\mu$. Thus the decay widths for the two-body nonleptonic
decays can be computed straightforward.

\section{Numerical Results and Discussions}

In this section, we present the numerical results.

In the numerical calculations, based on potential models the
parameters are chosen as follows:

$\lambda=0.24$ GeV$^2$, $\alpha=0.06$ GeV, $\Lambda_{QCD}=0.18$
GeV, $a=e=2.7183$, $V_0=-0.93$ GeV, $V_{bc}=0.04$\cite{data},
$m_1=1.846$GeV, $m_2=5.243$GeV.

With this set of parameters, we obtain the masses:
\begin{center}
$M_{B_c}=6.33$ GeV,  $\;\; M'=3.50$ GeV,
\end{center}
and corresponding radial wave-functions of $B_c$ meson and
$P$-wave charmonium $\chi_c, h_c$ numerically. Here in the present
evaluations, we only carry out the lowest order ones without
considering the splitting caused by $L-S$ and $S-S$ couplings, in
which all the bound states ${^3}P_J (J=0,1,2)$ and $^1P_1$ are
degenerated.

To see the behaviors of the universal function $\xi_1(t_m-t)$ and
$\xi_2(t_m-t)$ i.e. the two overlapping integrations of the wave
functions of initial and final states, we plot them explicitly in
Fig.1, where $t_m=(M-M')^2, t=(P-P')^2$.

The lepton energy spectra for the decays $B_c\rightarrow
\chi_c+e(\mu)+\nu$, for which the mass of charged lepton can be
ignored, are shown in Fig.2, and those for the decays
$B_c\rightarrow \chi_c+\tau+\nu$, for which the mass of charged
lepton $\tau$ cannot be ignored, are shown in Fig.3, where
$|\vec{p}_{\ell}|$ is the momentum of lepton. The difference
between Fig.2 and Fig.3 is due to the sizable mass of
$\tau$-lepton. For the semileptonic decays, we put the
corresponding widths in Table I.

As for the non-leptonic two-body decays $B_c \rightarrow
\chi_c(h_c)+h$, we only evaluate some typical channels, whose
widths are relatively larger, and put results in Table II. In the
numerical calculations, we have chosen $a_1=c_1$ and $a_2=c_2$,
i.e., $\kappa=0$, and $c_1$ and $c_2$ are computed at the energy
scale of $m_b$. The values of the decay constants:
$f_{\pi^+}=0.131$ GeV, $f_{\rho^+}=0.208$ GeV, $f_{a_1}=0.229$
GeV, $f_{K^+}=0.159$ GeV, $f_{K^{*+}}=0.214$ GeV, $f_{D_s}=0.213$
GeV, $f_{D^*_s}=0.242$ GeV, $f_{D^+}=0.209$ GeV,
$f_{D^{*+}}=0.237$ GeV are adopted by fitting decays of $B$ and
$D$ mesons.

If comparing the results in Table 1 with the decays of $B_c$ to
$S$-wave charmonium states $J/\psi$ and $\eta_c$ e.g.
$\Gamma(B_c\to J/\psi+l+\nu)\sim 25\cdot
10^{-15}$GeV\cite{cchd,dec}, one can realize the semileptonic
decays of $B_c$ to the $P$-wave charmonium states in magnitude are
about tenth of the decay $B_c\to J/\psi+l+\nu_l$.  As for the
two-body nonleptonic decays, due to the difference in momentum
recoil and the fact that the recoil momentum is fixed in a given
specific decay, the $P$-wave decay $B_c\to \chi_c(h_c)+h$ can be
greater than twentieth of the one, $B_c\to J/\psi(\eta_c)+h$, to
an $S$-wave state.

The first observation of $B_c$ by CDF group is through the
semileptonic decay $B_c\to J/\psi+l+\nu_l$, hence, we can conclude
that most of the decays concerned here are accessible in Run-II of
Tevatron and in LHC, especially, when the particular detector for
$B$ physics BTeV and LHCB at the two colliders are concerned. It
is because that Tevatron and LHC will have more than 20 time
events of $B_c$ meson than Run-I and have much better detectors.

Since the decays $B_c\to \chi_c[^3P_{1,2}]+l+\nu_l$ have such a
quite sizable branching ratio, so the cascade decays i.e. the
decays with an according one of the radiative decays
$\chi_c[^3P_{1,2}] \to J/\psi+\gamma$ followed may affect the
observation through the semileptonic decays $B_c\to
J/\psi+l+\nu_l$ as done by CDF group substantially, especially,
when the efficiency of detecting a photons for the detector is not
great enough.

We also would like to point out here that with sizable branching
ratio, the decays $B_c\to h_c+\l+\nu_l$ and/or $B_c\to h_c+h$
potentially can open a fresh `window' to observe the charmonium
state $h_c[^1P_1]$, especially, to note that the charmonium state
$h_c[^1P_1]$ has not been well-established experimentally yet.

%\noindent
\vspace{5mm}
%\newpage
\noindent
{\Large\bf Acknowledgement}
\vspace{5mm}

\noindent
This work was supported in part by National Natural
Science Foundation of China. The authors would like to thank J.-P.
Ma for valuable discussions. They also would like to thank G.T.
Bodwin for useful discussions.

\appendix
\section{}

In this appendix, we present the form factors and formulas for
$\alpha$, $\beta_{++}$ and $\gamma$ which are required in the
calculations on the exclusive semileptonic decays of $B_c$ to
$X_{c\bar c}$, which denotes one of $^1P_1$, $^3P_0$, $^3P_1$ and
$^3P_2$ states as indicated precisely in each case below.

For convenience, we introduce the parameters below:

$$\omega_{20}\equiv \omega'_{2}\frac{p\cdot p'}{MM'}\;, $$

$$ \omega_{10}\equiv \sqrt{\omega^{2}_{20}-m^2_2+m^2_1}\;,$$

$$nep={\sqrt {\frac{(p\cdot p')^2}{M'^2}-M^2}}\;.$$

\subsection{$B_c$ Meson to Charmonium $h_c[^1P_1]$}

The matrix elements for the vector and axial currents:

$$ <X(p', \epsilon)\vert V_{\mu}\vert B_c(p)>\equiv
r\epsilon^{*}_{\mu}+ s_{+}(\epsilon^{*}\cdot
p)(p+p')_{\mu}+s_{-}(\epsilon^{*}\cdot p)(p-p')_{\mu} $$

$$<X(p', \epsilon)\vert A_{\mu}\vert B_c(p)>\equiv iv
\epsilon_{\mu\nu\rho
\sigma}\epsilon^{*\nu}(p+p')^{\rho}(p-p')^{\sigma}. $$

Where

$$r=\frac{(m'_1-m_2)(m_1+\omega_{10}-m_2-\omega_{20})
\xi_2}{8m'_1\omega_{10}\omega_{20}} $$

\begin{equation}
-\frac{(m'_1+m_2)(m_1+\omega_{10}+m_2+\omega_{20}) \xi_2(p\cdot
p')}{8M'Mm'_1\omega_{10}\omega_{20}}\;, \\[2mm]
\end{equation}

$$s_{+}=\frac{m_2[M(m_2+\omega_{20}-m_1-\omega_{10})-M'(m_1+\omega_{10}+
m_2+\omega_{20})]\xi_2}{8M'M^2\omega_{10}{\omega_{20}}^2}
$$

$$+\frac{m_2[M(m_2+\omega_{20}-m_1-\omega_{10})-M'(m_1+\omega_{10}+
m_2+\omega_{20})]\xi_1}{8M'M\omega_{10}\omega_{20}nep} $$

$$+\frac{m_2[M(m_2\omega_{20}+\omega_{20}^2-m_1\omega_{20}-\omega_{10}^2)
-M'(m_1\omega_{20}-\omega_{10}^2+
m_2\omega_{20}+\omega_{20}^2)]\xi_2}{8M'M^2\omega_{10}^3\omega_{20}}
$$

\begin{equation}
 +\frac{(m'_1+m_2)(m_1+\omega_{10}+
m_2+\omega_{20})\xi_2}{16M'Mm'_1\omega_{10}\omega_{20}}-
\frac{(M'-M)\xi_1}{8M'M^2\omega_{10}}\;,
\end{equation}

$$s_{-}=\frac{m_2[-M(m_2+\omega_{20}-m_1-\omega_{10})-M'(m_1+\omega_{10}+
m_2+\omega_{20})]\xi_2}{8M'M^2\omega_{10}{\omega_{20}}^2} $$

$$+\frac{m_2[-M(m_2+\omega_{20}-m_1-\omega_{10})-M'(m_1+\omega_{10}+
m_2+\omega_{20})]\xi_1}{8M'M\omega_{10}\omega_{20}nep} $$

$$+\frac{m_2[-M(m_2\omega_{20}+\omega_{20}^2-m_1\omega_{20}-\omega_{10}^2)
-M'(m_1\omega_{20}-\omega_{10}^2+
m_2\omega_{20}+\omega_{20}^2)]\xi_2}{8M'M^2\omega_{10}^3\omega_{20}}
$$

\begin{equation}
-\frac{(m'_1+m_2)(m_1+\omega_{10}+
m_2+\omega_{20})\xi_2}{16M'Mm'_1\omega_{10}\omega_{20}}+
\frac{(M'-M)\xi_1}{8M'M^2\omega_{10}}\;,
\end{equation}

\begin{equation}
v=-\frac{(m'_1+m_2)(m_1+m_2+\omega_{10}+\omega_{20})
\xi_2}{16M'Mm'_1\omega_{10}\omega_{20}}\;.
\end{equation}

The dependence of $\alpha$, $\beta_{++}$ and $\gamma$ on the above
form factors:

\begin{equation}
\alpha=r^2+4M^2\stackrel{\rightarrow}{p'}^2v^2\;,
\end{equation}

\begin{center}

\begin{equation}
\beta_{++}=\frac{r^2}{4M'^2}-M^2yv^2+\frac{1}{2}\left[
\frac{M^2}{M'^2}(1-y)-1\right]rs_{+}+M^2\frac{\stackrel{\rightarrow}
{p'}^2}{M'^2}s^{2}_{+}\;,
\end{equation}

$$
\beta_{+-}=-\frac{r^2}{4M'^2}+(M^2-M'^2)v^2+\frac{1}{4}\left[
-\frac{M^2}{M'^2}(1-y)-3\right]rs_{+}$$
\begin{equation}
+\frac{1}{4}\left[ \frac{M^2}{M'^2}(1-y)-1\right]rs_{-}+
M^2\frac{\stackrel{\rightarrow}{p'}^2}{M'^2}s_{+}s_{-}\;,
\end{equation}

\begin{equation}
\beta_{-+}=\beta_{+-}\;,
\end{equation}

\begin{equation}
\beta_{--}=\frac{r^2}{4M'^2}+\left[M^2y-2(M^2+M'^2)\right]v^2+\frac{1}{2}
\left[-\frac{M^2}{M'^2}(1-y)-3\right]rs_{-}+
M^2\frac{\stackrel{\rightarrow} {p'}^2}{M'^2}s^2_{-}\;,
\end{equation}

\begin{equation}
\gamma=2rv\;.
\end{equation}
\end{center}

\subsection{$B_c$ Meson to Charmonium $\chi_c[^3P_0]$}

The matrix element for the vector current vanishes in the present
decay.

The matrix element for the axial current:

$$ <X(p')\vert A_{\mu}\vert B_c(p)>\equiv
u_{+}(p+p')_{\mu}+u_{-}(p-p')_{\mu}. $$

Where

$$u_{+}=\frac{[M(\omega_{10}+m_1+\omega_{20}+m_2)\xi_1+
M'(\omega_{20}+m_2-\omega_{10}-m_1)]m_2\xi_1}
{8\sqrt{3}M'\omega_{10}\omega_{20}nep} $$

$$-\frac{M'(\omega_{10}+m_1+\omega_{20}+m_2)\xi_1+
M(\omega_{20}+m_2-\omega_{10}-m_1)\xi_1}
{8\sqrt{3}M'\omega_{10}nep} $$

$$ +\frac{3\xi_2[M'(m'_1+m_2)(\omega_{10}+m_1+\omega_{20}+m_2)+
M(m'_1-m_2)(\omega_{20}+m_2-\omega_{10}-m_1)]}
{16\sqrt{3}M'Mm'_1\omega_{10}\omega_{20}} $$

$$ +\frac{\xi_2m_2[M'(-\omega_{10}-m_1+\omega_{20}+m_2)+
M(\omega_{20}+m_2+\omega_{10}+m_1)]}
{8\sqrt{3}M'M\omega_{10}\omega_{20}^2} $$

$$ +\frac{\xi_2[M'm_2(m_2\omega_{20}+\omega_{20}^2-
m_1\omega_{20}-\omega_{10}^2)+M\omega_{20}(m_2\omega_{20}+\omega_{20}^2+
m_1\omega_{20}-\omega_{10}^2)]}{8\sqrt{3}M'M\omega^{3}_{10}\omega_{20}}
$$

$$ +\frac{\xi_2[-M'(m_2\omega_{20}+\omega_{20}^2+
m_1\omega_{20}-\omega_{10}^2)+M(-m_2\omega_{20}-\omega_{20}^2+
m_1\omega_{20}+\omega_{10}^2)]}{8\sqrt{3}M'M\omega^{3}_{10}} $$

\begin{equation}
+\frac{\xi_2(M'-M)(m_2+\omega_{20})}{8\sqrt{3}M'Mm_2\omega_{10}}\;,\\[2mm]
\end{equation}

$$u_{-}=\frac{[-M(\omega_{10}+m_1+\omega_{20}+m_2)\xi_1+
M'(\omega_{20}+m_2-\omega_{10}-m_1)]m_2\xi_1}
{8\sqrt{3}M'\omega_{10}\omega_{20}nep} $$

$$-\frac{M'(\omega_{10}+m_1+\omega_{20}+m_2)\xi_1-
M(\omega_{20}+m_2-\omega_{10}-m_1)\xi_1}
{8\sqrt{3}M'\omega_{10}nep} $$

$$ +\frac{3\xi_2[M'(m'_1+m_2)(\omega_{10}+m_1+\omega_{20}+m_2)-
M(m'_1-m_2)(\omega_{20}+m_2-\omega_{10}-m_1)]}
{16\sqrt{3}M'Mm'_1\omega_{10}\omega_{20}} $$

$$ +\frac{\xi_2m_2[M'(-\omega_{10}-m_1+\omega_{20}+m_2)-
M(\omega_{20}+m_2+\omega_{10}+m_1)]}
{8\sqrt{3}M'M\omega_{10}\omega_{20}^2} $$

$$ +\frac{\xi_2[M'm_2(m_2\omega_{20}+\omega_{20}^2-
m_1\omega_{20}-\omega_{10}^2)-M\omega_{20}(m_2\omega_{20}+\omega_{20}^2+
m_1\omega_{20}-\omega_{10}^2)]}{8\sqrt{3}M'M\omega^{3}_{10}\omega_{20}}
$$

$$ +\frac{\xi_2[-M'(m_2\omega_{20}+\omega_{20}^2+
m_1\omega_{20}-\omega_{10}^2)-M(-m_2\omega_{20}-\omega_{20}^2+
m_1\omega_{20}+\omega_{10}^2)]}{8\sqrt{3}M'M\omega^{3}_{10}} $$

\begin{equation}
+\frac{\xi_2(M'+M)(m_2+\omega_{20})}{8\sqrt{3}M'Mm_2\omega_{10}}\;.%\\[2mm]
\end{equation}

The dependence of $\alpha$, $\beta_{++}$ and $\gamma$ on the above
form factors:

\begin{center}
\begin{equation}
\alpha=0,
\end{equation}
$$\beta_{++}=u^{2}_{+},\;\; \beta_{+-}=u_{+}u_{-}\;,$$
\begin{equation}
\beta_{-+}=u_{-}u_{+}, \;\;\; \beta_{--}=u^{2}_{-}\;,
\end{equation}
\begin{equation}
\gamma=0\;.%\\[2mm]
\end{equation}
\end{center}

\subsection{$B_c$ Meson to Charmonium $\chi_c[^3P_1]$}

The matrix elements for the vector and axial current currents:

$$ <X(p', \epsilon)\vert V_{\mu}\vert B_c(p)>\equiv
l\epsilon^{*}_{\mu}+ c_{+}(\epsilon^{*}\cdot
p)(p+p')_{\mu}+c_{-}(\epsilon^{*}\cdot p)(p-p')_{\mu}, $$

$$ <X(p', \epsilon)\vert A_{\mu}\vert B_c(p)>\equiv iq
\epsilon_{\mu\nu\rho\sigma}\epsilon^{*\nu}(p+p')^{\rho}(p-p')^{\sigma}.
$$

Where

$$l=\frac{(m_1+\omega_{10}+m_2+\omega_{20})\xi_1[(p\cdot
p')^2-M^2M'^2]m_2} {4\sqrt{2}MM'^2nep\omega_{10}\omega_{20}} $$

$$ -\frac{(m_1+\omega_{10}+m_2+\omega_{20})\xi_2(p\cdot p')}
{2\sqrt{2}MM'\omega_{10}\omega_{20}}
-\frac{(m_1+\omega_{10}-m_2-\omega_{20})\xi_2(p\cdot p')}
{4\sqrt{2}Mm'_1m_2\omega_{10}} $$

$$ +\frac{(m_1+\omega_{10}-m_2-\omega_{20})M'\xi_2}
{4\sqrt{2}m'_1\omega_{10}\omega_{20}} -\frac{[M'^2M^2-(p\cdot
p')^2] \xi_2}{4\sqrt{2}M'^2M^2\omega_{10}} $$

\begin{equation}
+\frac{\xi_2m_2[(p\cdot p')^2-M^2M'^2]}
{4\sqrt{2}M^2M'^2\omega_{10}\omega_{20}}
\left[\frac{(\omega_{10}+m_1+\omega_{20}+m_2)}{\omega_{20}}+
\frac{(m_1\omega_{20}+m_2\omega_{20}+\omega_{20}^2-
\omega_{10}^2)}{\omega_{10}^2} \right]\;,\\[2mm]
\end{equation}

$$c_{+}=\frac{(m_1+\omega_{10}+m_2+\omega_{20})\xi_1(M'^2-p\cdot
p')m_2}
{8\sqrt{2}MM'^2nep\omega_{10}\omega_{20}}+\frac{(M'^2-p\cdot p')
\xi_2}{8\sqrt{2}M'^2M^2\omega_{10}} $$

$$ +\frac{(m_1+\omega_{10}+m_2+\omega_{20})\xi_2}
{4\sqrt{2}MM'\omega_{10}\omega_{20}}
+\frac{(m_1+\omega_{10}-m_2-\omega_{20})\xi_2}
{8\sqrt{2}Mm'_1m_2\omega_{10}} $$ $$ +\frac{\xi_2m_2(M'^2-p\cdot
p')} {8\sqrt{2}M^2M'^2\omega_{10}\omega_{20}}
\left[\frac{(\omega_{10}+m_1+\omega_{20}+m_2)}{\omega_{20}}+
\frac{(m_1\omega_{20}+m_2\omega_{20}+\omega_{20}^2-
\omega_{10}^2)}{\omega_{10}^2} \right] $$

$$ c_{-}=\frac{(m_1+\omega_{10}+m_2+\omega_{20})\xi_1(M'^2+p\cdot
p')m_2}
{8\sqrt{2}MM'^2nep\omega_{10}\omega_{20}}+\frac{(M'^2+p\cdot p')
\xi_2}{8\sqrt{2}M'^2M^2\omega_{10}} $$

$$ -\frac{(m_1+\omega_{10}+m_2+\omega_{20})\xi_2}
{4\sqrt{2}MM'\omega_{10}\omega_{20}}
-\frac{(m_1+\omega_{10}-m_2-\omega_{20})\xi_2}
{8\sqrt{2}Mm'_1m_2\omega_{10}} $$

$$ +\frac{\xi_2m_2(M'^2+p\cdot p')}
{8\sqrt{2}M^2M'^2\omega_{10}\omega_{20}}
\left[\frac{(\omega_{10}+m_1+\omega_{20}+m_2)}{\omega_{20}}+
\frac{(m_1\omega_{20}+m_2\omega_{20}+\omega_{20}^2-
\omega_{10}^2)}{\omega_{10}^2} \right] $$

$$q=\frac{(\omega_{10}+m_1+\omega_{20}+m_2)\xi_1}{8\sqrt{2}
M'\omega_{10}nep}-\frac{m_2(-\omega_{10}-m_1+\omega_{20}+m_2)\xi_1}{8\sqrt{2}
M'\omega_{10}\omega_{20}nep} $$

$$+\frac{(\omega_{20}^2+m_1\omega_{20}+m_2\omega_{20}
-\omega_{10}^2)\xi_2}{8\sqrt{2}M'M\omega_{10}^3}
+\frac{(m_2+\omega_{20})\xi_2}{8\sqrt{2}M'Mm_2\omega_{10}} $$

$$ -\frac{\xi_2m_2} {8\sqrt{2}MM'\omega_{10}\omega_{20}}
\left[\frac{(-\omega_{10}-m_1+\omega_{20}+m_2)}{\omega_{20}}+
\frac{(-m_1\omega_{20}+m_2\omega_{20}+\omega_{20}^2-
\omega_{10}^2)}{\omega_{10}^2} \right]\;.$$

\vspace{2mm}

The dependence of $\alpha$, $\beta_{++}$ and $\gamma$ on the above
form factors:

\begin{center}
\begin{equation}
\alpha=l^2+4M^2\stackrel{\rightarrow}{p'}^2q^2,
\end{equation}\begin{equation}
\beta_{++}=\frac{l^2}{4M'^2}-M^2yq^2+\frac{1}{2}\left[
\frac{M^2}{M'^2}(1-y)-1\right]lc_{+}+M^2\frac{\stackrel{\rightarrow}
{p'}^2}{M'^2}c^{2}_{+},
\end{equation}
$$\beta_{+-}=-\frac{l^2}{4M'^2}+(M^2-M'^2)q^2+\frac{1}{4}\left[
-\frac{M^2}{M'^2}(1-y)-3\right]lc_{+}$$
\begin{equation}+\frac{1}{4}\left[
\frac{M^2}{M'^2}(1-y)-1\right]lc_{-}+M^2\frac{\stackrel{\rightarrow}
{p'}^2}{M'^2}c_{+}c_{-},
\end{equation}
\begin{equation}
\beta_{-+}=\beta_{+-}
\end{equation}
\begin{equation}
\beta_{--}=\frac{l^2}{4M'^2}+\left[M^2y-2(M^2+M'^2)\right]q^2+\frac{1}{2}\left[
-\frac{M^2}{M'^2}(1-y)-3\right]lc_{-}+M^2\frac{\stackrel{\rightarrow}
{p'}^2}{M'^2}c^2_{-}\;,
\end{equation}
\begin{equation}
\gamma=2lq\;.%\\[2mm]
\end{equation}
\end{center}

\subsection{$B_c$ Meson to Charmonium $\chi_c[^3P_2]$}

The matrix element of the vector and axial currents:

$$ <X(p', \epsilon)\vert V_{\mu}\vert B_c(p)>\equiv ih_{+-}
\epsilon_{\mu\nu\rho\sigma}\epsilon^{*\nu\alpha}
p_{\alpha}(p+p')^{\rho}(p-p')^{\sigma}, $$

$$ <X(p', \epsilon)\vert A_{\mu}\vert B_c(p)>\equiv
k\epsilon^{*}_{\mu\nu}p^{\nu}+
b_{+}(\epsilon^{*}_{\rho\sigma}p^{\rho}p^{\sigma})(p+p')_{\mu}+b_{-}
(\epsilon^{*}_{\rho\sigma}p^{\rho}p^{\sigma})(p-p')_{\mu}. $$

Where

$$k=-\frac{(m_1+
\omega_{10}+m_2+\omega_{20})\xi_1}{4\omega_{10}nep}
+\frac{m_2(-m_1-\omega_{10}+m_2+\omega_{20})\xi_1}
{4\omega_{10}\omega_{20}nep}$$

$$+\frac{(\omega_{10}^{2}-m_1\omega_{20}-
m_2\omega_{20}-\omega^{2}_{20})\xi_2}{4M\omega^{3}_{10}}-
\frac{(m_2+\omega_{20})\xi_2}{4Mm_2\omega_{10}} $$

$$ +\frac{\xi_2m_2} {4M\omega_{10}\omega_{20}}
\left[\frac{(-\omega_{10}-m_1+\omega_{20}+m_2)}{\omega_{20}}+
\frac{(-m_1\omega_{20}+m_2\omega_{20}+\omega_{20}^2-
\omega_{10}^2)}{\omega_{10}^2} \right] $$

$$b_{+}=\frac{m_2(m_1+\omega_{10}+m_2+\omega_{20})\xi_1}
{8M'M\omega_{10}\omega_{20}nep} +\frac{\xi_2}{8M'M^2\omega_{10}}$$

$$ +\frac{\xi_2m_2} {8M^2M'\omega_{10}\omega_{20}}
\left[\frac{(\omega_{10}+m_1+\omega_{20}+m_2)}{\omega_{20}}+
\frac{(m_1\omega_{20}+m_2\omega_{20}+\omega_{20}^2-
\omega_{10}^2)}{\omega_{10}^2} \right] $$

$$b_{-}=-\frac{m_2(m_1+\omega_{10}+m_2+\omega_{20})\xi_1}
{8M'M\omega_{10}\omega_{20}nep} -\frac{\xi_2}{8M'M^2\omega_{10}}$$

$$ -\frac{\xi_2m_2} {8M^2M'\omega_{10}\omega_{20}}
\left[\frac{(\omega_{10}+m_1+\omega_{20}+m_2)}{\omega_{20}}+
\frac{(m_1\omega_{20}+m_2\omega_{20}+\omega_{20}^2-
\omega_{10}^2)}{\omega_{10}^2} \right] $$

$$h_{+-}=-\frac{m_2(m_1+\omega_{10}+m_2+\omega_{20})\xi_1}
{8M'M\omega_{10}\omega_{20}nep} -\frac{\xi_2}{8M'M^2\omega_{10}}$$

$$ -\frac{\xi_2m_2} {8M^2M'\omega_{10}\omega_{20}}
\left[\frac{(\omega_{10}+m_1+\omega_{20}+m_2)}{\omega_{20}}+
\frac{(m_1\omega_{20}+m_2\omega_{20}+\omega_{20}^2-
\omega_{10}^2)}{\omega_{10}^2} \right]. $$

\vspace{2mm}

The dependence of $\alpha$, $\beta_{++}$ and $\gamma$ on the above
form factors:

\begin{center}
\begin{equation}
\alpha=\frac{M^2\stackrel{\rightarrow}{p'}^2}{2M'^2}
(k^2+4M^2\stackrel{\rightarrow}{p'}^2h^2),
\end{equation}
$$\beta_{++}=-\frac{yM^4\stackrel{\rightarrow}{p'}^2}{2M'^2}h^2+
\frac{M^2k^2}{24M'^2}\left(y+\frac{4\stackrel{\rightarrow}{p'}^2}
{M'^2}\right)+\frac{2b^{2}_{+}}{3}\frac{M^4\stackrel{\rightarrow}
{p'}^4}{M'^4}$$
\begin{equation}+
\frac{M^2\stackrel{\rightarrow}{p'}^2kb_{+}}{3M'^2}
\left[\frac{M^2}{M'^2}(1-y)-1\right],
\end{equation}
$$\beta_{+-}=\frac{M^2\stackrel{\rightarrow}{p'}^2}{2M'^2}h^2(M^2-M'^2)+
\frac{k^2}{24}\left(1-\frac{M^2}{M'^2}-\frac{4M^2\stackrel{\rightarrow}{p'}^2}
{M'^4}\right)+\frac{2b_{+}b_{-}}{3}\frac{M^4\stackrel{\rightarrow}
{p'}^4}{M'^4}$$
\begin{equation}+
\frac{M^2\stackrel{\rightarrow}{p'}^2kb_{+}}{6M'^2}
\left[-\frac{M^2}{M'^2}(1-y)-3\right]-
\frac{M^2\stackrel{\rightarrow}{p'}^2kb_{-}}{6M'^2}
\left[-\frac{M^2}{M'^2}(1-y)+1\right],
\end{equation}
\begin{equation}\beta_{-+}=\beta_{+-}\end{equation}
$$\beta_{--}=-\frac{M^2\stackrel{\rightarrow}{p'}^2}{2M'^2}h^2
\left[2(M^2+M'^2)-M^2y\right]
+\frac{2b^{2}_{-}}{3}\frac{M^4\stackrel{\rightarrow}
{p'}^4}{M'^4}$$
\begin{equation}
+\frac{k^2}{24}\left(2+\frac{M^2}{M'^2}(2-y)+\frac{4M^2\stackrel{\rightarrow}{p'}^2}
{M'^4}\right)-
\frac{M^2\stackrel{\rightarrow}{p'}^2kb_{-}}{3M'^2}
\left[\frac{M^2}{M'^2}(1-y)+3\right],
\end{equation}
\begin{equation}
\gamma=\frac{M^2\stackrel{\rightarrow}{p'}^2kh}{M'^2}.
\end{equation}
\end{center}

%\newpage
\begin{table}
\begin{center}
\caption{The semileptonic decay widths (in the unit $10^{-15}$
GeV)}

\begin{tabular}{|c|c|c|c|c|}\hline
&$\Gamma(B_{c}{\longrightarrow}{^1P_1}
{\ell}{ {\nu}}_{\ell})$&$\Gamma(B_{c}{\longrightarrow}{^3P_0}
{\ell}{ {\nu}}_{\ell})$&$\Gamma(B_{c}{\longrightarrow}{^3P_1}
{\ell}{ {\nu}}_{\ell})$&$\Gamma(B_{c}{\longrightarrow}{^3P_2}
{\ell}{ {\nu}}_{\ell})$\\ \hline
%$e(\mu)$&2.022&1.338&1.801&2.014\\\hline
%$\tau $&0.295&0.216&0.299&0.316\\\hline
$e(\mu)$&2.509&1.686&2.206&2.732\\\hline
$\tau $&0.356&0.249&0.346&0.422\\\hline
\end{tabular}
\end{center}
%\end{table}

%\newpage
%\begin{table}
\begin{center}
\caption{Two-body non-leptonic $B_c^+$ decay widths in unit
$10^{-15}$ GeV}

\vspace*{0.5 cm}
\begin{tabular}{|c|c|c||c|c|c|}
~~Channel~~&$\Gamma $  ~ &$\Gamma(a_1=1.132)$&
~~Channel~~&$\Gamma $ ~ &$\Gamma(a_1=1.132)$ \\
$^1P_1 \pi^+$ & $a_1^2 ~ 0.569$ &$0.729$~~ &
$^1P_1 \rho$   & $a_1^2 ~ 1.40$  &$1.79$~~ \\
$^3P_0 \pi^+$ & $a_1^2 ~ 0.317$ &$0.407$~~ &
$^3P_0 \rho$   & $a_1^2 ~ 0.806$  &$1.03$~~ \\
$^3P_1 \pi^+$ & $a_1^2 ~ 0.0815$ &$0.104$~~ &
$^3P_1 \rho$   & $a_1^2 ~ 0.331$  &$0.425$~~ \\
$^3P_2 \pi^+$ & $a_1^2 ~ 0.277$ &$0.355$~~ &
$^3P_2 \rho$   & $a_1^2 ~ 0.579$  &$0.742$~~ \\
&&&&&\\
$^1P_1 A_1$ & $a_1^2 ~ 1.71$ &$2.19$~~ &
$^1P_1 K^+$   & $a_1^2 ~ 4.26\times 10^{-3}$  &$5.46 \times 10^{-3}$~~ \\
$^3P_0 A_1$ & $a_1^2 ~ 1.03$ &$1.33$~~ &
$^3P_0 K^+$   & $a_1^2 ~ 2.35\times 10^{-3}$  &$3.02 \times 10^{-3}$~~ \\
$^3P_1 A_1$ & $a_1^2 ~ 0.671$ &$0.859$~~ &
$^3P_1 K^+$   & $a_1^2 ~ 0.583\times 10^{-3}$  &$0.747 \times 10^{-3}$~~ \\
$^3P_2 A_1$& $a_1^2 ~ 1.05$ &$1.34$~~ &
$^3P_2 K^+$& $a_1^2 ~ 1.99\times 10^{-3}$&$2.56 \times 10^{-3}$~~\\
&&&&&\\
$^1P_1 K^{*}$ & $a_1^2 ~ 7.63\times 10^{-3}$ &$9.78 \times 10^{-3}$~~ &
$^1P_1 D_s$   & $a_1^2 ~ 2.32$  &$2.98$~~ \\
$^3P_0 K^{*}$ & $a_1^2 ~ 4.43\times 10^{-3}$ &$5.68 \times 10^{-3}$~~ &
$^3P_0 D_s$   & $a_1^2 ~ 1.18$  &$1.51$~~ \\
$^3P_1 K^{*}$ & $a_1^2 ~ 2.05\times 10^{-3}$ &$2.63 \times 10^{-3}$~~ &
$^3P_1 D_s$   & $a_1^2 ~ 0.149$  &$0.191$~~ \\
$^3P_2 K^{*}$ & $a_1^2 ~ 3.48\times 10^{-3}$ &$4.47 \times 10^{-3}$~~ &
$^3P_2 D_s$   & $a_1^2 ~ 0.507$  &$0.650$~~ \\
&&&&&\\
$^1P_1 D^*_s$ & $a_1^2 ~ 1.99$ &$2.56$~~ &
$^1P_1 D^+$   & $a_1^2 ~ 0.0868$  &$0.111$~~ \\
$^3P_0 D^*_s$ & $a_1^2 ~ 1.48$ &$1.89$~~ &
$^3P_0 D^+$   & $a_1^2 ~ 0.0443$  &$0.0568$~~ \\
$^3P_1 D^*_s$ & $a_1^2 ~ 2.21$ &$2.83$~~ &
$^3P_1 D^+$   & $a_1^2 ~ 0.00610$  &$0.00782$~~ \\
$^3P_2 D^*_s$ & $a_1^2 ~ 2.68$ &$3.44$~~ &
$^3P_2 D^+$   & $a_1^2 ~ 0.0209$  &$0.0267$~~ \\
&&&&&\\
$^1P_1 D^{*+}$ & $a_1^2 ~ 0.0788$ &$0.101$~~ & & &\\
$^3P_0 D^{*+}$ & $a_1^2 ~ 0.0567$ &$0.0726$~~ & & &\\
$^3P_1 D^{*+}$ & $a_1^2 ~ 0.0767$ &$0.0983$~~ & & &\\
$^3P_2 D^{*+}$ & $a_1^2 ~ 0.0972$ &$0.124$~~ & & &~~
 \end{tabular}
\end{center}
\label{tab3}
\end{table}

\begin{figure}\begin{center}
   \epsfig{file=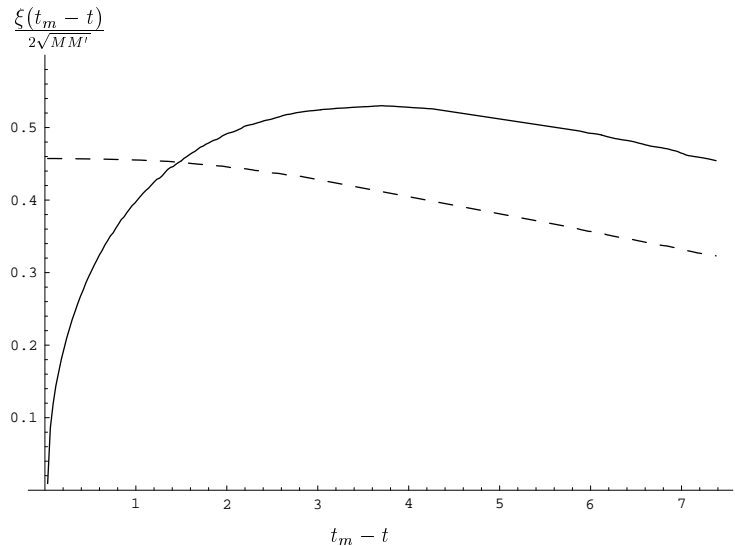, bbllx=160pt,bblly=350pt,bburx=550pt,bbury=660pt,
width=10cm,angle=0} \caption{The universal functions $\xi_1$ and
$\xi_2$ vs. $t_m-t$. They are the overlapping-integrations of the
the wave functions for $\chi_c(h_c)$ and $B_c$ with the definition
as in Eq.(38). The solid line is of $\xi_1$, the dashed one is of
$\xi_2$.}
\end{center}
\end{figure}

\begin{figure}\begin{center}
   \epsfig{file=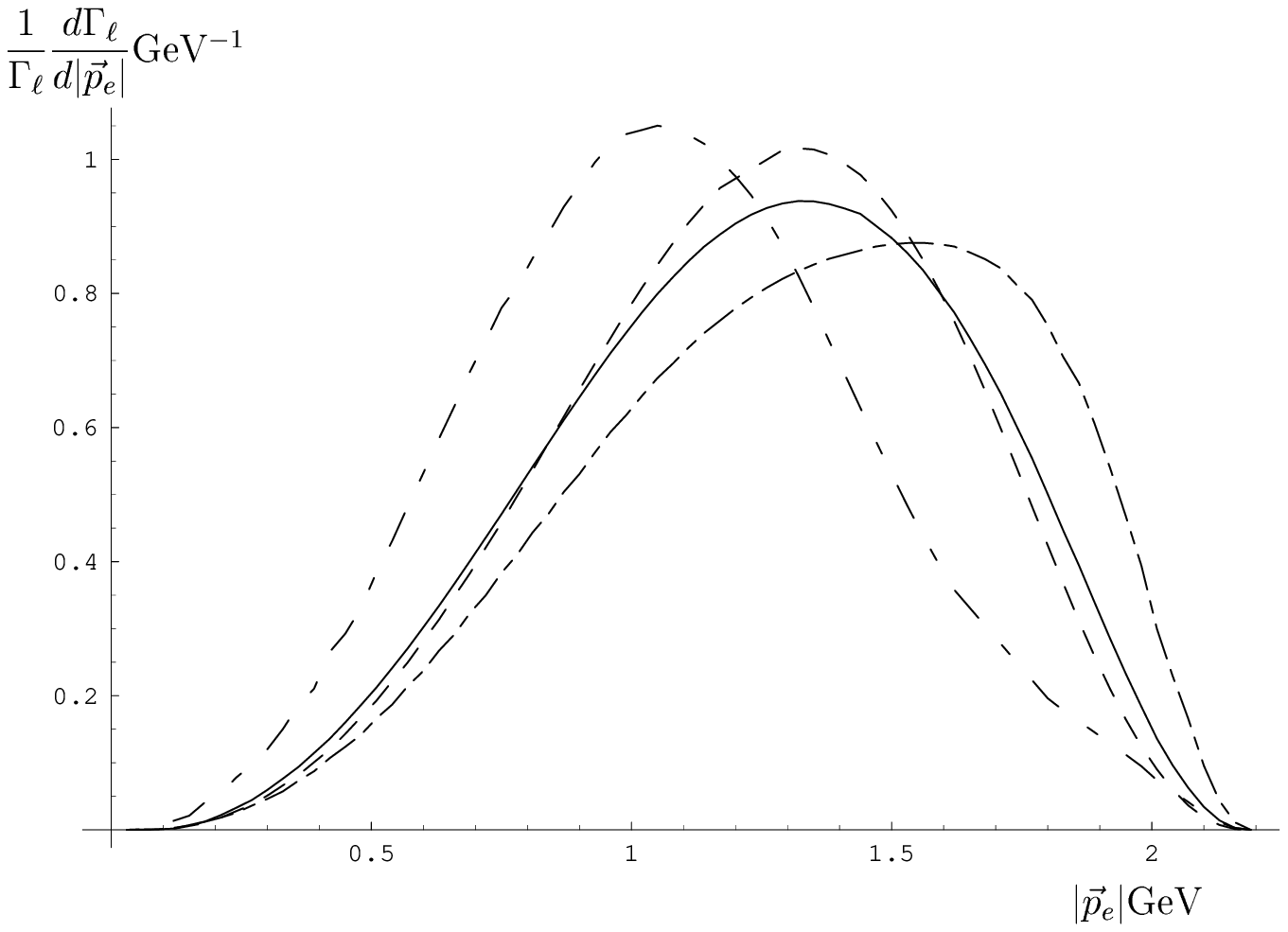, bbllx=160pt,bblly=350pt,bburx=550pt,bbury=660pt,
width=10cm,angle=0} \caption{{The energy spectrum of the charged
lepton for the decays $B_c\rightarrow \chi_c+e(\mu)+\nu$, where
the solid line is the result of $h_c[^1P_1]$ state,
dotted-blank-dashed line is of $\chi_c[^3P_0]$, dashed line is of
$\chi_c[^3P_1]$, dotted-dashed line is $\chi_c[^3P_2]$.}}
\end{center}
\end{figure}

\begin{figure}\begin{center}
   \epsfig{file=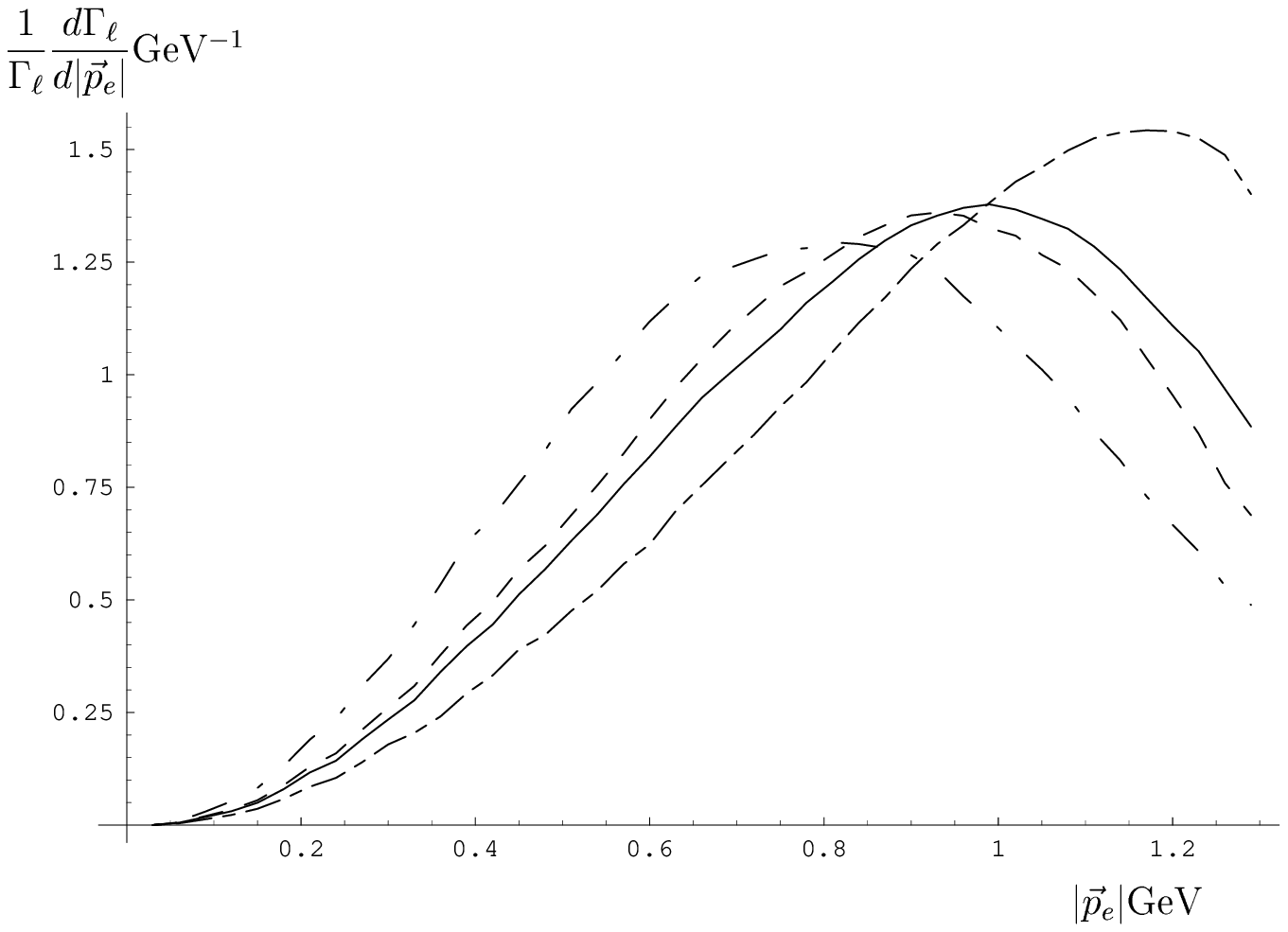, bbllx=160pt,bblly=350pt,bburx=550pt,bbury=660pt,
width=10cm,angle=0} \caption{{The energy spectrum of the charged
lepton for the decays $B_c\rightarrow \chi_c+\tau+\nu_{\tau}$,
where the solid line is the result of $h_c[^1P_1]$ state,
dotted-blank-dashed line is of $\chi_c[^3P_0]$, dashed line is of
$\chi_c[^3P_1]$, dotted-dashed line is $\chi_c[^3P_2]$.}}
\end{center}
\end{figure}

\end{document}